\documentclass[galaxies,article,accept,moreauthors,10pt,a4paper]{mdpi}

\firstpage{1}
\articlenumber{x}
\doinum{10.3390/------}
\pubvolume{5}
\pubyear{2017}
\copyrightyear{2017}
\externaleditor{{Academic Editor: Tomotsugu Goto}}
\history{Received: 31 July 2017; Accepted: 18 September 2017; Published: date}

\usepackage[flushleft]{threeparttable}
\usepackage{mathabx}
\usepackage{upgreek}
 \theoremstyle{mdpi}
 \newcounter{thm}
 \newcounter{ex}
 \newcounter{re}
 \theoremstyle{mdpidefinition}

\Title{{Upper Limits to Magnetic Fields in the Outskirts of~Galaxies}}

\Author{{Ericson L\'opez} *,
{Jairo Armijos-Abenda\~no},
Mario Llerena and Franklin Ald\'as}
\AuthorNames{Firstname Lastname, Firstname Lastname and Firstname Lastname}

\address[1]{%
{Observatorio} 
Astron\'omico de Quito, Escuela Polit\'ecnica Nacional, Av. Gran Colombia S/N, {Quito 170403}, {Ecuador};
 jairo.armijos@epn.edu.ec (J.A.-A.); mario.llerena01@epn.edu.ec (M.L.); franklin.aldas@epn.edu.ec (F.A.)}

\corres{\hangafter=1 \hangindent=1.05em \hspace{-0.82em} Correspondence: ericsson.lopez@epn.edu.ec}

\abstract{Based on CO(2-1) public data, we study the monoxide oxygen gas excitation conditions and the magnetic field strength of four spiral galaxies. For the galaxy outskirts, we found kinetic temperatures in the range of $\lesssim$35--38 K, CO column densities $\lesssim 10^{15}$--$10^{16}$ cm$^{-2}$, and H$_2$ masses $\lesssim 4\times 10^6$--$6\times10^8$ M$_\odot$. An H$_2$ density $\lesssim 10^3$ cm$^{-3}$ is suitable to explain the 2$\sigma$ upper limits of the CO(2-1) line intensity. We constrain the magnetic field strength for our sample of spiral galaxies and their outskirts by using their masses and H$_2$ densities to evaluate a simplified magneto-hydrodynamic equation. Our estimations provide values for the magnetic field strength on the order of $\lesssim$6--31 {$\upmu$G}.}


\begin{document}




\section{Introduction}

In this paper, we focus our attention on the study of the {magnetic field strength in the outskirts of four spiral galaxies,} following a different approach {to those commonly based on Faraday rotation}, dust polarization, synchrotron emission, etc. To constrain the magnetic field strength of spiral galaxies, we will follow {the Dotson method, described in Section 4.4 of \cite{Dotson}}; i.e., approaching the magneto-hydrodynamic force equation to derive a simple expression to estimate the upper limit of the magnetic field.

\subsection*{Magnetic Field Constraint}
As mentioned above, we use the Dotson method \cite{Dotson}  to constrain the magnetic field,  which is mainly based on the following relation:
\begin{equation}\label{eq1}
B<3.23\times 10^{-8}\left(\dfrac{R}{\mathrm{pc}}\right)^{0.5}\left(\dfrac{n}{ \mathrm{cm^{-3}} }\right)^{0.5}\left(\dfrac{M}{\mathrm{M_{\odot}}}\right)^{0.5}\left(\dfrac{r}{\mathrm{pc}}\right)^{-1}
\end{equation}
\noindent
where $R$ is the radius of the magnetic field lines, $n$ is the molecular hydrogen gas density, $M$ and $r$ are the total mass and radius of the source, respectively. So, the magneto-hydrodynamic force can be used to estimate the magnetic field strength (Equation \eqref{eq1}) when there is available information about the shape of the field lines (see \cite{Dotson}).
On the other hand, to estimate the density $n$ and mass $M$ of the source, we use the carbon monoxide emission as a tracer of the molecular gas H$_2$ (\cite{Neininger}). This is because H$_2$ is invisible in the cold interstellar medium (around 10--20 K), so its distribution and motion must be inferred from observations of minor constituents of the clouds, such as carbon monoxide and dust~(\cite{Neininger}). Carbon monoxide is the most abundant molecule after H$_2$, CO~is easily excited, and~the emission of CO(1-0) at 2.6 mm is ubiquitous in the Galaxy (\cite{Neininger}). So, CO it is a good tracer for molecular~hydrogen.

\section{Carbon Monoxide Data}\label{sec:carbon}

\begin{figure}[H]
\centering
\includegraphics[width=10cm]{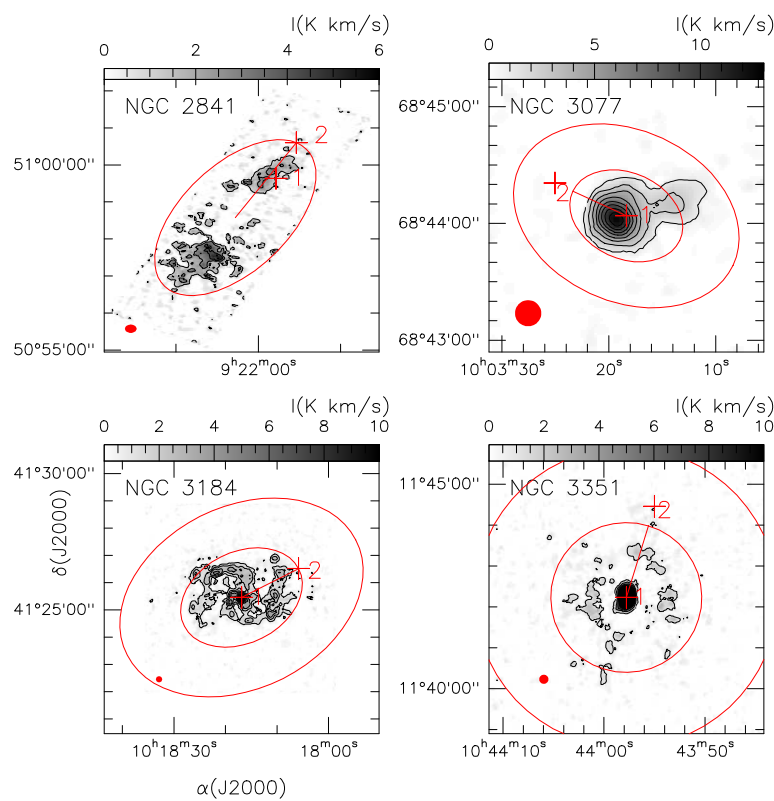}\vspace{-1mm}
\caption{{CO(2-1)}
integrated intensity maps of our galaxy sample. The red crosses show positions used to extract the spectra shown in Figure \ref{fig2}. The red ellipses indicate the regions used to measure CO(2-1) luminosities (see text). The tiny filled red ellipse represents the IRAM telescope beam (13 arcsec at the CO(2-1) frequency transition of 230.538 GHz). For NGC 2841, we did not measure the CO(2-1) luminosity in the galaxy outskirts, as in this particular case was not observed.}\label{fig1}
\end{figure}

To carry out this study, we used public CO(2-1) data, first published by  \cite{Leroy}, data obtained with the {IRAM 30 m} 
telescope\footnote{\url{http://www.iram-institute.org/EN/30-meter-telescope.php}}
located in Spain. At  the CO(2-1) transition frequency ({230.538} GHz),
the~IRAM telescope has a spatial resolution of $13$~arcsec. From the available data we selected a sample of four nearby spiral galaxies, whose morphology and positions are listed in Table \ref{table1}. On the other hand, in~Figure~\ref{fig1}, CO(2-1) integrated intensity maps of each galaxy in our sample are shown.

\vspace{-1mm}\begin{table}[H]
\caption{Galaxy sample morphology and positions.}\label{table1}\vspace{-4mm}
\begin{center}
\begin{threeparttable}
\small
\centering
\begin{tabular}{ccccc}
\toprule
\multirow{2}{*}{\textbf{Galaxy Name}}	& \textbf{{RA}
~\tnote{1}}	& \textbf{{DEC}
~\tnote{1}} & \multirow{2}{*}{\textbf{Morphology} \tnote{1}} & \textbf{Distance \tnote{1}}\\
                &  {\bf (hh:mm:ss.s)} & {\bf (hh:mm:ss.s)} & & {\bf (Mpc)}\\
\midrule
NGC 2841		& 09:22:02.7	& +50:58:35.3 & SAa C & 14.6\\
NGC 3077		& 10:03:19.1	& +68:44:02.2 & S0 C  &  3.8\\
NGC 3184        & 10:18:17.0    & +41:25:27.8 & SAc C & 11.3\\
NGC 3351        & 10:43:57.7    & +11:42:13.0 & SBb C & 10.5\\
\bottomrule
\end{tabular}
\begin{tablenotes}
\centering\footnotesize {\item[1] Information taken from the SIMBAD Astronomical Database.}
\end{tablenotes}
\end{threeparttable}
\end{center}
\end{table}\unskip\vspace{-2mm}

\section{Data Treatment}\unskip
\subsection{Spectra Selection}\label{sel_spectra}

As was mentioned in the introduction, in order to constrain the magnetic field strengths for the galaxies in our sample, we use the method applied by \cite{Dotson}, which is based on Equation \eqref{eq1}. Therefore, for every source, the parameters $R$, $n$, $M$, and $r$ should be determined or assumed. We estimate $n$, $M$, and $r$ from CO(2-1) observations (see Section \ref{results}), while $R$ is assumed based on the results of previous {works} \cite{Knapik, Beck, Chyzy, Fletcher}.

\begin{figure}[H]
\centering
\includegraphics[width=10cm]{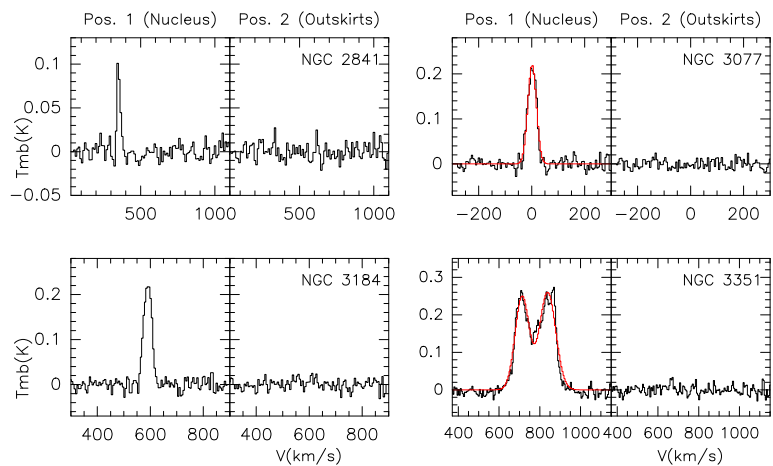}\vspace{-1mm}
\caption{Spectra extracted from positions 1 and 2 indicated in Figure \ref{fig1}. As an example, for NGC 3077 and NGC 3351 galaxies, we show the Gaussian fits (indicated with red curves) applied to the CO(2-1)~lines.}\label{fig2}
\end{figure}

To derive $n$ and $M$ from the CO(2-1) data we have selected the spectra by choosing two positions; one located in the nucleus of the galaxy (hereinafter the position 1) and the second located on the outskirts of the galaxy (hereinafter the position 2). CO(2-1) line emission is not detected towards the NGC 2841 nucleus; therefore, as an exception, its position 1 spectrum corresponds to a position displaced 79 arcsec from the galaxy nucleus. The inner ellipse is taken as the one that holds as much as possible the integrated intensity of CO (2-1) radiation emitted by the galaxy. The center of this ellipse defines the nucleus (position 1). Starting from the nuclei, in steps of 20 arcsec, we take CO(2-1) spectrum along the major axis of the inner ellipse. Position 2 is defined as the more contiguous positions to the galaxy nucleus, placed along the galaxy major-axis, where the CO(2-1) line emission is no longer detected above 2$\sigma$. We assume that position 2 is representative of the lowest boundary of the galactic outskirts which are well traced by the HI emission (\cite{Sofue}). In Figure \ref{fig1}, positions 1 and 2 are indicated by red crosses. For all galaxies in our sample, the spectra extracted for both positions, within the 13 arcsec resolution  of the IRAM telescope, are shown in Figure \ref{fig2}. In our study, the choice of position 2 depends on two aspects: (1) the IRAM telescope beam size and (2) the step of 20 arcsec used to find one of the nearest position to the galaxy nucleus, along the major-axis, where CO(2-1) emission is no longer detected.

\subsection{Gaussian Fitting and CO Luminosities}

Gaussian fits to the CO(2-1) lines (indicated in Figure \ref{fig2}) have been performed. Then, the large velocity gradient (LVG) modeling (\cite{Van}) is employed to estimate the gas density ($n$), based on the CO(2-1) line width ($\Delta V$) and the line intensity ($I$). For {position}
~1 (nucleus) we derived the values of these parameters, which are presented in Table \ref{table2}. For {position}
~2 (outskirts), we estimated {2$\sigma$ upper limits} for the CO(2-1) line intensity, which are also given in Table \ref{table2}. This table includes the radial velocity ($V$) and the average $\overline{\Delta V}$ obtained from spectra extracted along the major-axis of the galaxy disk.

To estimate CO integrated luminosities ($L_{CO}$) for the galaxies in our sample, we used the inner  ellipse for the nuclear region (the $L_{CO}$ will be used later to estimate H$_2$ masses). As seen in Figure \ref{fig1}, and as it was indicated before, this ellipse is defined to enclose almost all the CO(2-1) emission arising from the galaxy disk. In turn, to estimate a $2\sigma$ upper limit on the $L_{CO}$ for the galaxy outskirts, we use a ringlike region with axes equal to twice the sizes of the inner ellipse. The derived $L_{CO}$ values for both regions of each galaxy are given in Table \ref{table3}.


\begin{table}[H]
\centering
\caption{CO(2-1) line properties derived for our galaxy sample.}\label{table2}
\begin{tabular}{cccccc}
\toprule
\multirow{2}{*}{\bf Galaxy Name\vspace{-5pt}}	& \multirow{2}{*}{\bf Region\vspace{-5pt}} & \boldmath{$\Delta V$}{\bf (}\boldmath{$\sigma$}{\bf )}	&  \boldmath{$V$}{\bf (}\boldmath{$\sigma$}{\bf )}&\boldmath{$I$\,$^{(2)}$} & \boldmath{$\overline{\Delta V}$\,$^{(3)}$} \\\cmidrule{3-6}
& & {\bf km s}\boldmath{$^{-1}$} & {\bf km s}\boldmath{$^{-1}$} &{\bf K} & {\bf km s}\boldmath{$^{-1}$}	\\
\midrule
\multirow{3}{*}{NGC 2841\vspace{-5pt}}& Position displaced 79 arcsec&\multirow{2}{*}{25.2 (2.0)}&	\multirow{2}{*}{352.0}	&\multirow{2}{*}{0.1}&	\multirow{2}{*}{23.4}\\\vspace{0.3mm}
        & from nucleus $^{(1)}$   &    &    &    &    \\
        & outskirts	&... &... &$\lesssim$0.03& ...\\\midrule
\multirow{2}{*}{NGC 3077}& nucleus   &38.2 (1.1) &1.8 (0.5) &0.2& 40.2	\\
	    & outskirts &... &...& $\lesssim$0.02 &... \\\midrule
\multirow{2}{*}{NGC 3184}& nucleus	&35.4 (1.0) &590.3 (0.5) &0.2	&25.6\\
	    & outskirts &... &...&	$\lesssim$0.03&	...\\\midrule
\multirow{2}{*}{NGC 3351}& nucleus	&80.9 (2.3) &713.1 (0.9)& 0.3 & 45.9 \\
        & outskirts &... &...& $\lesssim$0.03& ...\\
\bottomrule
\end{tabular}
\begin{tabular}{@{}c@{}}
\multicolumn{1}{p{\textwidth -.88in}}{\footnotesize $^{(1)}$As CO(2-1) line emission is not detected at the galaxy nucleus, this position is displaced {79 arcsec}
from the galaxy nucleus (see Figure \ref{fig1}); $^{(2)}$ For galaxy outskirts, this parameter is a 2$\sigma$ upper limit based on the rms noise; $^{(3)}$Average value obtained from several spectra extracted along the major-axis of the galaxy disk.}
\end{tabular}

\end{table}\unskip

\begin{table}[H]
\centering
\caption{Parameters derived for the galaxy sample.}\label{table3}
\tablesize{\footnotesize}
{
\centering
\begin{tabular}{ccccccc}
\toprule
\multirow{2}{*}{\bf Galaxy Name\vspace{-5pt}}& \multirow{2}{*}{\bf Region\vspace{-5pt}} & \boldmath{$L_{CO}$\,$^{(2)}$} &  \boldmath{$r$\,$^{(3)}$} &  \boldmath{$M_{H_2}$\,$^{(4)}$} & {\bf (} \boldmath{{$M_{H_2}+M_{HI}$}}{\bf )} \boldmath{$^{(5)}$} &
 \boldmath{$B$}\\\cmidrule{3-7}
& & \boldmath{$\times$}{\bf 10} \boldmath{$^6$} {\bf K km s} \boldmath{$^{-1}$} {\bf pc} \boldmath{$^2$} & {\bf kpc} & \boldmath{$\times$}{\bf 10} \boldmath{$^7$} {\bf M} \boldmath{{$_{\odot}$}}
&  \boldmath{$\times$}{\bf 10} \boldmath{$^8$} {\bf M} \boldmath{$_\odot$}	&  \boldmath{$\upmu$}{\bf G}\\ \midrule
\multirow{2}{*}{NGC 2841} &disk         &2.0& 8.5   &103.0&89.6&$\lesssim$31\\
         &outskirts $^{(1)}$&...& ...  &...  &... &...\\\midrule
\multirow{2}{*}{NGC 3077}&disk     &2.3      &0.4   &1.3      &0.4      &$\lesssim$6\\
        &outskirts&$\lesssim$0.7&0.8   &$\lesssim$0.4&$\lesssim$0.1&$\lesssim$7\\\midrule
\multirow{2}{*}{NGC 3184}&disk     &188.0    &6.6   &103.0 &36.8&$\lesssim$14\\
        &outskirts&$\lesssim$81.1&13.2&$\lesssim$44.6&$\lesssim$15.9&$\lesssim$19\\\midrule
\multirow{2}{*}{NGC 3351}&disk&226.0&7.1 &124.0&32.1&$\lesssim$11\\
        &outskirts&$\lesssim$109.0&14.2&$\lesssim$59.8&$\lesssim$15.5&$\lesssim$15\\
\bottomrule
\end{tabular}}
\begin{tabular}{@{}c@{}}
\multicolumn{1}{p{\textwidth -.88in}}{\footnotesize $^{(1)}$ For NGC 2841 we were not able to estimate $L_{CO}$ luminosity and other relative parameters in the outskirts, as this region was not observed; $^{(2),(4),(5)}$ For galaxy outskirts, these parameters are a 2$\sigma$ upper limit; $^{(3)}$~This~radius traces the boundary of the galaxy disk region, where the CO(2-1) line emission is still detectable above 2$\sigma$ (see Section \ref{sel_spectra}).}
\end{tabular}

\end{table}

\section{Results\label{results}}\unskip

\subsection{Galaxy Mass}
The molecular hydrogen mass ($M_{H_2}$) for the galaxy disk and its outskirts is derived using the  the equation given {by} \cite{Leroy}:
\begin{equation}\label{eq2}
{\dfrac{M_{H_2}}{\mathrm{M_{\odot}}}=5.5\,\dfrac{R_{21}}{0.8}\left(\dfrac{L_{CO}}{\mathrm{K \, km \, s^{-1}\,  pc^2}}\right)}
\end{equation}
where $R_{21}$ is the CO(2-1)/CO(1-0) line intensity ratio (equal to 0.8) and $L_{CO}$ is the CO luminosity. In~this equation, a CO(1-0)/H$_2$ conversion factor of $2\times 10^{20}$ cm$^{-1}$ (K km s$^{-1}$)$^{-1}$ is adopted \cite{Leroy}. In~Table~\ref{table3}, the $M_{H_2}$ values are listed  for both galaxy regions. We found that the H$_2$ outskirts masses derived for our galaxies  sample vary within $\lesssim 4\times 10^6$--$1\times 10^9$ M$_\odot$. The total mass ($M_{H_2}+M_{HI}$) is also presented in the same table. The atomic hydrogen mass has been estimated by using the $M_{H_2}$/$M_{HI}$ ratio given in \cite{Leroy}.

\subsection{H$_2$ Density, CO Column Density, and Kinetic Temperature}

As was mentioned above, the H$_2$ density is derived for the nucleus and outskirts positions in the galaxies of our sample by using the LVG approach. We fit the CO(2-1) line intensity or its limit and the average line width $\overline{\Delta V}$, while the CO column density ($N_{CO}$), H$_2$ molecular density ($n_{H_2}$), and~the kinetic temperature ($T_{kin}$) are considered as free parameters. The CO(1-0) line intensity is known based on the CO(2-1)/CO(1-0) line intensity ratio of about 0.8 derived by \cite{Leroy}. Note that at {position~2 (i.e., outskirt positions)}
, we do not detect CO(2-1) emission and we have derived 2$\sigma$ upper limits for these line intensities. We found that a $n_{H_2}$ density of about 10$^3$ cm$^{-3}$ is suitable to fit the CO(2-1) line intensity for the four galaxies in our sample. For positions 1 and 2, the $N_{CO}$ and $T_{kin}$ that provide the best fits to the CO(2-1) line intensity (or its limit) are given in Table \ref{table4}. For the galaxy outskirts, we found $T_{kin} \lesssim$ 35--38 K and $N_{CO} \lesssim 10^{15}$--$10^{16}$ cm$^{-2}$.These findings tell us that the molecular gas in galaxy outskirts is relatively cold. For the nucleus positions, the $T_{kin}$ are found within 35--38 K and the $N_{CO}$ within $\sim 10^{15}$--$10^{16}$ cm$^{-2}$. So, the column density N$_{CO}$ exhibits the greatest changes, whereas the kinetic temperature T$_{kin}$ is relatively constant.  The physical parameters presented in Table \ref{table4}  seem to be similar for both nuclei and outskirts, but  in the case of the nuclei the given values are accurate, while for the outskirts they correspond  to the 2$\sigma$ upper limits. 

\begin{table}[H]
\caption{Physical parameters derived for our galaxy sample.}\label{table4}\vspace{-4mm}
\begin{center}
\small 
\centering
\begin{tabular}{cccc}
\toprule
{\bf Component}&\boldmath{$n_{H_2}$}&\boldmath{$T_{kin}$}&\boldmath{$N_{CO}$}\\
\midrule
outskirts&$\lesssim 10^3$ cm$^{-3}$&$\lesssim$35--38 K& $\lesssim 10^{15}$--$10^{16}$ cm$^{-2}$\\
nucleus&$10^3$&35--38 K&$\sim 10^{15}$--$10^{16}$ cm$^{-2}$\\
\bottomrule
\end{tabular}
\end{center}
\end{table}

\subsection{Magnetic Fields in Galaxies and Their Outskirts}

As  mentioned previously, in order to constrain the magnetic field strength for a given galaxy and its outskirts, we use Equation \eqref{eq1} given by \cite{Dotson}. This equation includes the $n_{H_2}$, the mass $M$, and radius $r$ of the galaxy, and the radius of curvature $R$ of the magnetic field lines. $n_{H_2}$ and $M$ were derived in the previous sections, but the mass that we use in Equation \eqref{eq1} {refers}
~to the dust mass, which is obtained by the relation ($M_{H_2}+M_{HI}$)/100; i.e., assuming the typical dust-to-gas mass ratio of 0.01. In~this section, we assume that $R$ is equal to the radius $r$ of the studied regions. This assumption is crude, but it is based on several studies of spiral galaxies such as NGC 4736, M51, NGC~1097, and NGC~1365, which reveal magnetic field lines that extend as far as their galactic arms (\cite{Chyzy, Fletcher, Beck}). The~derived values for the magnetic field strength are listed in Table \ref{table3}. For the galaxy outskirts, we have considered the upper limits of $M$ and $n$ as fixed values in order to calculate the magnetic field strength. For~the NGC 2841 galaxy, no value was estimated for the magnetic field in its outskirts, because this region was not observed in this object. We found magnetic field magnitudes on the order of $\lesssim$6--31 $\upmu$G for the galaxies in our sample and their outskirts. These limits agree with those values of $\sim$20--60 $\upmu$G found in spiral galaxies (\cite{Knapik, Beck, Fletcher}).


\section{Conclusions}

In the present contribution, we have estimated the magnetic field strength in the galaxy nuclei and in the outskirts of a sample of four spiral galaxies. For that, we have used an approximate expression of the magnetohydrodynamics to find an upper limit for the magnetic field magnitudes. The~magnetic field strength lies within the range of $\lesssim$6--31 $\upmu$G, which is in good agreement with the values provided by other authors for spiral galaxies. A first good idea about the strength of the magnetic field can be obtained directly from the estimation of molecular hydrogen mass and radio of the source, without the necessity of a magnetohydrodynamical model or the use of a traditional technique like Faraday rotation or Zeeman line broadening. This is a rough estimation that works if the gas pressure is uniform and the viscosity is neglected. A better approach can be obtained by keeping more terms in the magnetohydrodynamics force equation \cite{Dotson} to impede gravitational collapse.

Moreover, instead of using the total hydrogen mass ($M_{H_2} + M_{HI}$) in our magnetic field calculations, the mass of the dust  has been considered as enough to get values in good agreement with the \mbox{$ \sim$20--60 $\upmu$G} observed in spiral galaxies (\cite {Knapik, Beck, Fletcher}). This fact suggests that  the dust is the main component that  influences the magnetic field strength better than the molecular gas.

On the other hand, for the galaxy outskirts we found a kinetic temperature $T_{kin}$ of about  \mbox{$\lesssim$35--38 K} and a column density $N_{CO}$ $\lesssim10^{15}$--$10^{16}$ cm$^{-2}$. These findings tell us that the molecular gas in galaxy outskirts is relatively cold. Moreover, the {$M_{H_2}$}
masses on the outskirts of the galaxy are in the range of $\lesssim 4\times 10^6$--$6\times 10^8$ $\mathrm{M_\odot}$, and a {$n_{H_2}$} density of $\lesssim$10$^3$ cm$^{-3}$ is suitable to explain the 2$\sigma$ upper limits to the CO(2-1) line intensity.
Then, it seems that  if the densities and temperatures are low in the outskirts it would result in a higher {$M_{H_2}$} mass at a given CO(2-1) luminosity if the outskirts volume is increasing. It is interesting that the values of both H$_2$ density and kinetic temperature are relatively similar in the nuclear region of the studied galaxies, but not the CO column density that varies by a factor of 10~along our sample.

{In future research, we plan}
~to go deeper in understanding the magnetic field structure in galaxy halos, studying more spiral galaxies and employing other molecular hydrogen tracers. Additionally,~understanding the variations of the field direction associated with the column density changes is part of our future work.










\vspace{6pt}\authorcontributions{E. L\'opez, \mbox{{J. Armijos-Abenda\~no}}, M. Llerena and F. Ald\'as performed the data analysis. E. L\'opez and \mbox{J. {Armijos-Abenda\~no}} wrote the manuscript. All authors contributed to the discussion and interpretation of the results.}

\conflictsofinterest{The authors declare no conflict of interest.}



\reftitle{References}
\bibliographystyle{mdpi}

\renewcommand\bibname{References}



\end{document}